\documentclass[conference]{IEEEtran}
\IEEEoverridecommandlockouts

\usepackage{cite}
\usepackage{amsmath,amssymb,amsfonts}
\usepackage{graphicx}
\usepackage{algorithm}
\usepackage{tabularx}
\usepackage{booktabs}
\usepackage{url}
\usepackage[noend]{algpseudocode}
\usepackage{mathtools}
\usepackage{subfigure}

\usepackage{graphicx}
\usepackage{textcomp}
\usepackage{xcolor}

\def\BibTeX{{\rm B\kern-.05em{\sc i\kern-.025em b}\kern-.08em
    T\kern-.1667em\lower.7ex\hbox{E}\kern-.125emX}}
\begin{document}

\title{CloudSimSC: A Toolkit for Modeling and Simulation of Serverless Computing Environments\\

}

\author{\IEEEauthorblockN{Anupama Mampage and Rajkumar Buyya}
\IEEEauthorblockA{\textit{Cloud Computing and Distributed Systems (CLOUDS) Laboratory} \\
\textit{School of Computing and Information Systems}\\
\textit{The University of Melbourne, Australia}  \\
Email: mampage@student.unimelb.edu.au, rbuyya@unimelb.edu.au
}

}

\maketitle

\begin{abstract}

Serverless computing is gaining traction as an attractive model for the deployment of a multitude of workloads in the cloud. Designing and building effective resource management solutions for any computing environment requires extensive long term testing, experimentation and analysis of the achieved performance metrics. Utilizing real test beds and serverless platforms for such experimentation work is often times not possible due to resource, time and cost constraints. Thus, employing simulators to model these environments is key to overcoming the challenge of examining the viability of such novel ideas for resource management. Existing simulation software developed for serverless environments lack generalizibility in terms of their architecture as well as the various aspects of resource management, where most are purely focused on modeling function performance under a specific platform architecture. In contrast, we have developed a serverless simulation model with induced flexibility in its architecture as well as the key resource management aspects of function scheduling and scaling. Further, we incorporate techniques for easily deriving monitoring metrics required for evaluating any implemented solutions by users. Our work is presented as CloudSimSC, a modular extension to CloudSim which is a simulator tool extensively used for modeling cloud environments by the research community. We discuss the implemented features in our simulation tool using multiple use cases. 

\end{abstract}


\section{Introduction}
Serverless computing has gained much popularity as a deployment model for applications in the cloud during the past few years. Enterprise cloud users from numerous application domains have expressed their interest in adopting this novel computing model owing to the simplification it brings to the whole end to end process of acquiring and managing cloud resources for hosting their software components. Under this novel paradigm, the cloud vendor takes on the full responsibility of managing the servers on behalf of the users. This means that the entire array of operational tasks of provisioning the required resources, allocating them to running applications, scheduling the applications on the infrastructure, scaling the allocated resources to suit existing traffic levels and monitoring application performance is fully under the control of the cloud service provider. Recent statistics show that more than half of the customers of many industry giants in cloud computing such as AWS, Google and Azure, have adopted the serverless offerings of these companies. This evolving market of serverless users is estimated to be worth around \$8 billion at the moment.

Along with presenting a blackbox front from the perspective of the end users comes the heavy duty of handling all the operational tasks related to the application backends, to the service providers themselves. In addition, unlike the operational scenarios of leased out Virtual Machine (VM) infrastructure, serverless environments are of a multi-tenant nature. This means that numerous functions belonging to multiple users would reside in a shared space and consume the virtual resources simultaneously. Further, the cloud vendors are expected to manage the cloud resources for serverless applications with minimum input and intervention from the application owners, in order to better maintain the 'serverless' notion of infrastructure management. Accordingly, the service providers would initially have only a minimal understanding on the behavior of the deployed functions. This significantly complicates their task of managing the allocated cloud resources to the applications while satisfying their performance requirements as well as the budgetary restrictions of cloud infrastructure costs. 

Although the 'serverless' nature of application resource management relieves end users from many responsibilities, it also creates some additional complications not present under a conventional cloud computing model. Inability to recreate the execution environment for an application is one such major shortcoming, which hinders the performance as well as budgetary planning processes for the users. The design of individual functions and the structure of an application requires a sound understanding on the run time resource consumption and behaviour of the same, so that the optimum code-level decisions can be made. 

\begin{table*}[!t]
	\caption{A comparision with related serverless computing simulators}
	\label{table:relatedWork}
	\resizebox{\textwidth}{!}{\begin{tabular}{l c c c c c c c}
			\toprule
			\multicolumn{1}{c}{\textbf{Work}} &\multicolumn{2}{c}{\textbf{Architecture}} &\multicolumn{3}{c}{\textbf{Configurable Resource Management}}
			&\multicolumn{2}{c}{\textbf{Monitoring Perspective}}
			\\
			
			& \textbf{Single Request} &\textbf{Request Concurrency}&\textbf{Scheduling}&\textbf{Horizontal Scaling}&\textbf{Vertical Scaling}&\textbf{Application owner}&\textbf{Service Provider} \\
			\hline
			Mahmoudi et al. (2021) &\checkmark & &&&&\checkmark &\checkmark\\
			\hline
			Jeon et al. (2019)&\checkmark&&\checkmark&&&\checkmark&\\
			\hline
                Mastenbroek et al. (2021)&\checkmark&\checkmark&\checkmark&&&\checkmark&\\
			\hline
			CloudSimSC (Our work) &\checkmark&\checkmark&\checkmark&\checkmark&\checkmark&\checkmark&\checkmark\\
			\bottomrule
	\end{tabular}}
\end{table*}

The availability of a simulator platform replicating the practical serverless computing infrastructure is thus essential for the serverless providers to make better resource management decisions in a multi-tenant enviornment as well as for the end users to plan and design their applications for the most favorable end results. The very few existing simulator environments presented by researchers for serverless environments have a narrowed scope, limiting to specific architectures of a particular commercial platform and focusing mostly on performance modeling for a deployed function. In contrast, our simulator CloudSimSC, aims to provide a generalized serverless environment capable of following a flexible function execution flow customizable depending on the commercial/open source platform that needs to be replicated. Further, our environment is developed with a focus of allowing the incorporation of custom function scaling and scheduling policies, forming the dual perspective of both the serverless application and infrastructure owners. We develop this environment as an extension to a widely adopted simulator in cloud environments named, CloudSim \cite{calheiros2011cloudsim}. Thus, CloudSimSC could be used along with all the existing functionalities of the CloudSim base simulator.   

The major \textbf{contributions} of our work are listed below.
\begin{enumerate}

\item A generalized architecture for function execution following both the existing commercial and open-source serverless architectures, allowing the users to choose depending on the environment that needs to be replicated.

\item Functional components for request load balancing, function scheduling and resource scaling (horizontal and vertical)

\item Facilitates the introduction of custom load balancing, scheduling and scaling policies from the perspective of service providers. Custom scaling options for horizontal and vertical scaling of function instances are provided based on the chosen function execution flow.

\item Ability to derive monitoring metrics with regard to application performance, system throughput and the underlying resource consumption for infrastructure providers.

\end{enumerate}
With these developments built on CloudSim, CloudSimSC brings all the benefits of a simulation approach, wherein performance evaluation experiments can be conducted under repeatable simulation environments for multiple system configurations and applications and user QoS (Quality of Service) requirement scenarios. This developed serverless simulation environment has been instrumental in conducting all the performance evaluation experiments in our previous work \cite{mampage2021deadline}.

\section{Related Work}

A few attempts have been made to develop or extend existing simulators to recreate serverless computing environments. Mahmoudi et al. \cite{mahmoudi2021simfaas} present a simulator developed in python following the available public serverless computing platforms. Accordingly, they use only the scale-per-request architecture seen in commercial platforms such as AWS Lambda \cite{WhatisAW74:online}, Google Cloud platform \cite{GoogleCl97:online}, Azure Cloud Functions \cite{AzureFun24:online}, where a function instance would serve only a single request at a time. Their simulator identifies the different states of a function instance and the cold/warm starts of a function request. They also focus on modeling the performance of functions by determining metrics such as the average response time, probability of cold starts and the average cost of infrastructure measured in terms of the number of  machines in use. However they lack focus on the generalizability of function scheduling or scaling functionalities. 

Jeon et al. \cite{jeon2019cloudsim} develop an extension to cloudSim simulator introducing a geo-distributed serverless architecture including an edge network. They allow functions to define service level objectives (SLOs) but the included features do not necessarily follow the execution flow of existing serverless platforms in use, but rather presents a visionary architecture for serverless computing at the edge.

An open-source platform OpenDC is developed by Mastenbroek et al. \cite{mastenbroek2021opendc} for modeling and simulation of emerging cloud datacenter technologies. The presented usecases for datacenter simulation in this platform include serverless computing among many others such as machine learning and HPC-as-a-service infrastructure. Under the serverless implementation, they focus on different policies for resource allocation and scheduling for functions, in addition to the cost models for function resource consumption for the users. However, the focus on provider aspect of operations and generalizability of platform architecture along with the scaling of function resources is lacking. 

Table 1 summarizes the comparison among the existing serverless simulators with CloudSimSC in terms of the presented architecture, focus on function scheduling, horizontal and vertical scalability of deployed function instances, and the derivation of monitoring metrics useful for both the application owners and infrastructure owners. The architecture refers to whether a single function instance is able to accommodate multiple requests at a time. Based on this feature, routing of an incoming request, scheduling of a function instance on a VM and subsequently scaling up/down and in/out of function resources need to be handled considering different factors.

From Table 1, we can observe that most of the existing simulators lack generalizability in terms of the architecture for function execution since most of them only support the primary design specifications seen in commerial serverless platforms. Open-source platforms such as OpenFaas \cite{HomeOpen35:online}, Kubeless \cite{Kubeless69:online} and Fission  \cite{FissionF58:online} are built based on the Kubernetes \cite{Kubernet74:online} container orchestration tool at their core. Accordingly, they follow a slightly different execution pattern allowing multiple requests to be executed simultaneously on a single function instance, referred to as a pod. Our simulator allows the user to decide on which execution style they want to incorporate. The existing simulator designs also do not specifically focus on the configurability of new scheduling or scaling policies with ease whereas ours support easy inclusion of new scheduling policies, horizontal, and even vertical scaling policies depending on the usecase. Further, we introduce the provider perspective in metrics monitoring, specially with regard to the cost of maintaining the serverless infrastructure which is disregarded by many.

\section{Design of CloudSimSC and its Components}

\begin{figure*}[!t]
	\centering 
	\includegraphics[width=15cm, height=10cm]{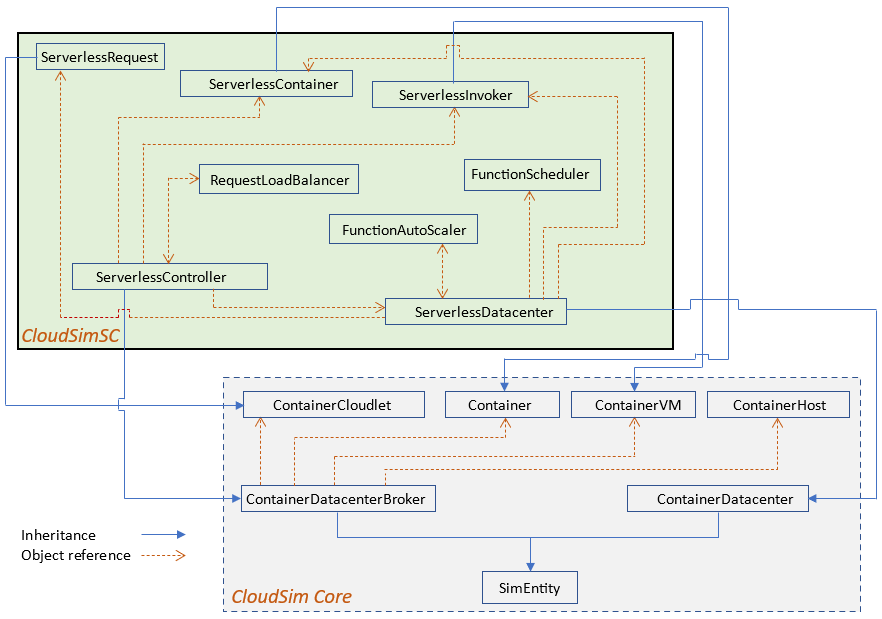}
	\caption{A class diagram of CloudSimSC as an extension of the existing CloudSim simulator}
	\label{fig:ClassDiagram}
\end{figure*}

To address the limitations seen in existing simulator designs, we introduce CloudSimSC, an extension to the CloudSim simulator which facilitate the execution of serverless functions. 

Fig. 1 shows how the features of the CloudSimSC extension are intergrated with the existing components of the CloudSim simulator. CloudSim is a highly modular framework extensively used by researchers for modeling, simulation and experimentation of various cloud application and resource provisioning scenarios in cloud computing environments over the years. CloudSim has the abstract class SimEntity at its core, representing the simulation entities that are able to send and process messages from and to other entities. All communications and actions in the simulator take place via events represented by SimEvent objects. Events are stored and executed sequentially as per their simulation time, by invoking their corresponding methods. The ContainerDatacenterBroker and ContainerDatacenter classes are derived from the SimEntity class. The ContainerDatacenter class represents the core hardware infrastructure maintained by cloud providers. This class in turn encapsulates the ContainerHost and ContainerVM classes which denote the physical and virtual machines for running cloud applications.

The core components of the CloudSim framework can be extended to simulate various distributed computing scenarios within the cloud computing paradigm. Thus we have developed the CloudSimSC framework extending CloudSim, which enables the inclusion of unique features of a serverless computing environment in order to easily simulate and model function executions. As seen in Fig. 1, we have inherited many of the core classes in CloudSim such as the ContainerDatacenterBroker, ContainerDatacenter, ContainerHost, ContainerVM, Container, and ContainerCloudlet and extended their features to support serverless executions. These components are explained in detail in the following subsections. Further, we illustrate the basic serverless system model followed in our design in Fig. 2, which helps in understanding these key functional elements.

\begin{figure}[!b]
	\centering 
	\includegraphics[width=\linewidth, height=6.5cm]{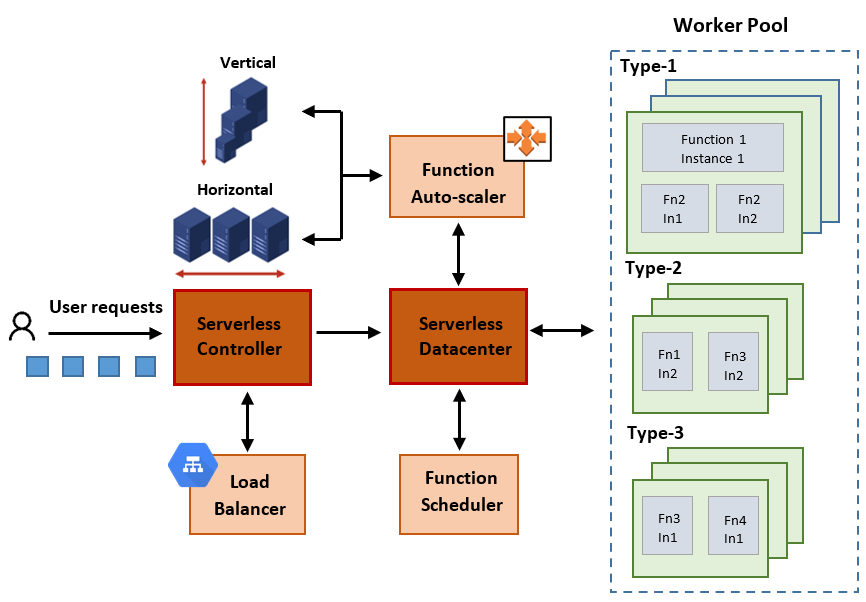}
	\caption{The system model of the serverless application execution environment}
	\label{fig:SystemModel}
\end{figure}

\subsection{ServerlessController}
ServerlessController class extends the ContDatacenterBroker class. This acts as the coordinating body which handles communication between the end users and the cloud service providers offering serverless computing services. It handles online negotiations for reserving required resources for meeting user demands to their satisfaction. The controller receives the external user requests and directs them to the load balancer for deciding its' execution environment depending on the user specified platform architecture. It also monitors and gathers the status of cluster resources represented by the ServerlessDatacenter class as well as the performance data of serverless workload executions. The gathered data could be used by clients for measuring the quality of service of the cloud providers and also for budgeting purposes. Cloud providers could use the same for evaluating their resource management techniques, which is crucial for the provider managed serverless computing paradigm. 

\subsection{RequestLoadBalancer}

\begin{algorithm}[!b]
	\caption{Load Balancer}\label{algo:loadbalancer}
	\begin{algorithmic}[1]
		\Procedure{LoadBalancing}{$r$}
            \State $v \leftarrow getVMsCreatedList()$
            \If {$scalePerRequest$}
            \State {$createNewContainer()$}            
            \Else
            \State $contTypeExists \leftarrow false$
            \State $contAllocated \leftarrow false$
            \For {$vm := v$}            
            \State $contList \leftarrow vm.getFunctionContainerMap(r.type)$
            \If {$contList != null$}
            \State $contTypeExists \leftarrow true$
            \EndIf
            \For {$c := contList$}
            \If{$resource_c^{avail} \geq resource_r^{req}$}
            \State $r.setContainer(c)$
            \State $contAllocated \leftarrow true$
            \State $submitRequest(r)$
            \State  \textbf{break}
            \EndIf
            \EndFor
            \If{$contAllocated = true$}
             \State  \textbf{break}
             \EndIf
             \EndFor
             \If{$contAllocated = false$}
             \If{$contTypeExists = false$}
             \State $pendingContList \leftarrow vm.getFunctionContainerMapPending(r.type)$
            \If {$pendingContList != null$}
            \State $contTypeExists \leftarrow true$
            \EndIf
             \EndIf                 
            \If{ $contTypeExists = true$}
             \State  $reScheduleRequest(r)$
             \Else
             \State $createNewContainer(r)$
             \EndIf
             \EndIf            
            \EndIf
		\EndProcedure
	\end{algorithmic}
\end{algorithm}

All the user requests are initially received and queued at the load balancer. The execution flow of a request from this point is dependent on the selected platform architecture by the end user. If the user intends to follow the commercial serverless computing architecture with a single request occupying a container at a time, they could choose to either create a new container for every new request (scale per request) or to select an existing idle container to schedule the incoming request, if available. If the user decides to follow a scale per request style, the new container creation request is submitted to the ServerlessController, who passes it on to the ServerlessDataCenter. Subsequently, the request is scheduled on the newly created container. If instead the user wishes to use any existing warm, idling containers that are available before creating new containers, containerIdling needs to be enabled which commands every container to be retained for a set time interval after usage. If the user plans to follow a serverless architecture more synonymous with the open-source platforms, they could opt to have request concurrency in function containers where multiple requests of the same function are accommodated by a function instance at a time. In this case, they can either choose to have containerIdling enabled or have a separate auto-scaling policy for maintaining a pool of ready instances. 

In using either of the platform architectures, whenever idle instances are available, a suitable policy for container selection need to be implemented under the selectContainer method, where the user requests are forwarded to an existing function instance with sufficient free resources. The default implementation has the First Fit (FF) logic where the first available matching instance is selected for request execution. If an active idle container is currently not available, the load balancer checks for any suitable containers pending to be created. If such pending containers exist, the request could wait for a set time interval for a scheduling retry when the container gets launched. If this is not an option, a new container gets created for the request execution. Algorithm \ref{algo:loadbalancer} illustrates this procedure, where $r$ denotes an incoming user request. 

\subsection{ServerlessDatacenter}
ServerlessDatacenter class encapsulates the hardware/virtual infrastructure of the serverless computing environment. It receives forwarded requests by the load balancer and handles their execution. The datacenter manages the operation of the bare metal host nodes, the virtual machines that run on their resources and also the separate sandbox environments created for request executions such as containers. The FunctionScheduler and FunctionAutoScaler are initialized as objects within the ServerlessDatacenter class. All cluster infrastructure resource metrics are also constantly gathered and stored within the datacenter.

\subsection{FunctionScheduler}

\begin{algorithm}[t]
	\caption{Function Auto-Scaler} \label{algo:resourcescaler}
	\begin{algorithmic}[1]
		\Procedure{ContainerScalingTrigger}{}
            \State $r_{L} \leftarrow getFunctionTypes()$
            \State $v \leftarrow getVMsCreatedList()$
            \For {$vm := v$}
            \State $contMap \leftarrow vm.getFunctionContainerMap()$
            
            \For {$r := r_{L}$}
            \For {$c := contMap(r)$}
            \State $fnDataMap \leftarrow c.getResourceData$
            \EndFor
            \EndFor
            \EndFor
            \State \textbf{return} $fnDataMap$
		\EndProcedure
            \Procedure{horizontalScaler}{$fnDataMap$, $r_{L}$}
            \For {$r := r_{L}$}
            \For {$c := fnDataMap(r)$}
            \State $d_{r} \leftarrow calculateDesiredReplicas(fnDataMap)$
            \State $n_{r} \leftarrow calculateNewReplicas(d_{r})$
            \If{$n_{r}$$ > $$ 0$}
            \For {$i = 1$ to $ n_{r}$}
            \State $createScaledContainer()$
            \EndFor
            \Else
            \For {$i = 1$ to $ n_{r}$}
            \State $destroyIdleContainers()$         
            \EndFor
            \EndIf
            \EndFor
            \EndFor
		\EndProcedure
            
	\end{algorithmic}
\end{algorithm}

The function scheduler is tasked with selecting a suitable VM for scheduling a newly created container. An object of this class is initialized with the creation of an instance of the datacenter, which identifies the allocation policy for containers on VMs. Since serverless systems are multi-tenant environments, functions of multiple users could reside on the same VM. These different applications would have varying resource requirements and thus the occurrence of resource pressure on VM resources is quite prevalent. Further, the sensitivity of these multitude of applications to various resource conditions too could have vast differences. Hence the strategy for selecting a VM to run a function instance should ideally take into account all these factors. 
The container to be scheduled could already be reserved for executing a particular user request or it could be a new container request triggered by the auto-scaler component. The container could also be an instance with the capacity to host multiple concurrent user requests or a single request at a time. Depending on the vCPU and memory requirements and availability in the cluster VMs, the function scheduling logic is to be incorporated under the findVmForContainer method. The default implementation consists of the RR, random and bin-packing policies.

\subsection{FunctionAutoScaler}
A distinct property of the serverless computing paradigm is its fine grained auto-scaling capabilities, which aims to follow workload traffic patterns closely and create only the resources that are needed, when they are needed. This adhoc creation of resources results in unprecedented delays in function execution. As such, different strategies are employed by commercial and open-source platforms to have ready-to-serve instances for incoming requests, without compromising too much on resource wastage. These techniques are either focused on retaining used instances for future reuse or on proactively creating new instances by predicting the load levels. Existing commercial serverless platforms mostly manage function auto-scaling by maintaining used containers for a certain time duration to serve any new requests. The open-source serverless platforms which are based on kubernetes follow threshold based horizontal scaling where the user can select to scale resources when the average resource utilization across the function instances or the requests per second for a function exceeds a set threshold.

Function scaler in CloudSimSC is capable of both horizontal and vertical scaling of function resources. Horizontal scaling is associated with increasing or decreasing the number of replicas of a function, i.e., scaling in or out. Vertical scaling on the other hand handles the increase and decrease of the resource capacities of a single instance of a function (container). Adjusting resource capacities in this way is useful for when the time needed for creating new resources with the horizontal scaler is not tolerable for an application. Further, this allows to improve resource utilization levels of the provider VM nodes. Thus using vertical scaling in conjunction with horizontal scaling increases the opportunity of achieving better resource efficiency for the cloud provider and better function performance for the end users. In CloudSimSC, when the scaling functionality is enabled, the function scaler periodically performs the scaling functionalities as per the implemented policies. The default implemented policy for triggering function scaling follows a threshold based logic. Whenever the average cpu utilization level of function instances of a particular type rises beyond the set threshold, the scaler is invoked. 

\subsubsection{Horizontal Scaler}
Once triggered, the horizontal scaler could follow any logic for creating new replicas, as required by different users. The default policy follows a simple logic of calculating the number of replicas required to bring the utilization level to the set threshold.

\subsubsection{Vertical Scaler}
For the vertical scaler, we define a set of cpu and memory increment levels that a function could refer to. Accordingly, for each function type we need to identify the cpu and memory incremental actions, that could be successfully performed considering resource restrictions in the host nodes and the requests that are already in execution in the existing instances. The default implementation includes the selection of a random scaling action from the viable options.

Algorithm \ref{algo:resourcescaler} presents the default auto-scaler logic inlcuding the threshold based horizontal scaling policy implementation.

\begin{figure}[!ht]
	\centering 
	\includegraphics[width=\linewidth, height=4cm]{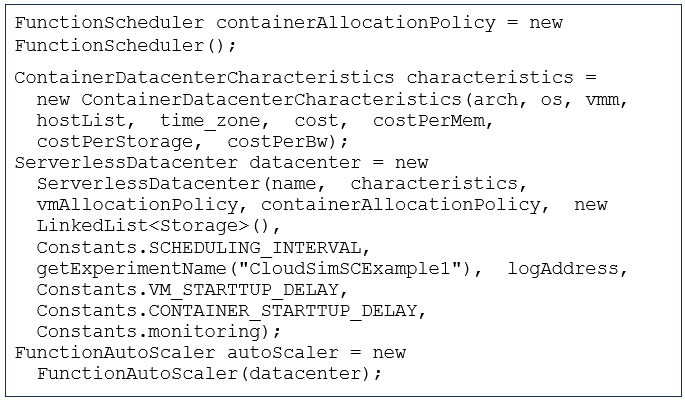}
	\caption{Sample code snippet for creating a ServerlessDatacenter}
	\label{fig:Code3}
\end{figure}

\begin{figure}[!ht]
	\centering 
	\includegraphics[width=\linewidth, height=4cm]{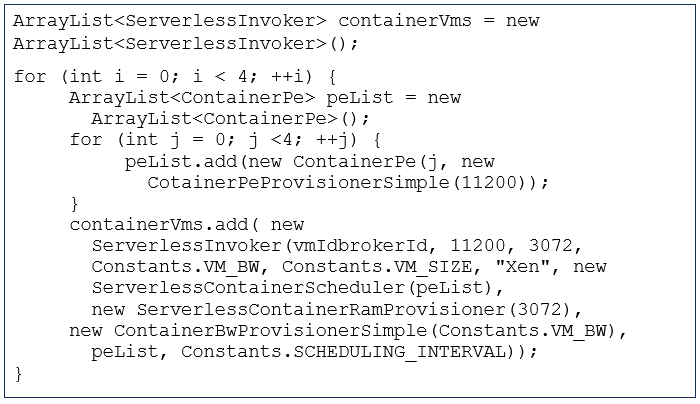}
	\caption{Sample code snippet for creating a vm cluster}
	\label{fig:Code4}
\end{figure}

\section{Sample Simulation using CloudSimSC}
In this section we present the step by step execution of a sample simulation scenario in CloudSimSC. Simulation parameters corresponding to a particular scenario are to be configured in the Constants class file.

\textit{Step1}.  Initialize the CloudSim core operations using the CloudSim.init() function.

\textit{Step2}. Create a ServerlessController instance. The controller handles all communication to and from the user and the serverless cluster environment. 

\textit{Step3}. Create a ServerlessDatacenter instance along with a set of host nodes within it. Objects of the FunctionScheduler component which identifies the containerAllocationPolicy and also the FunctionAutoScaler are also initialized together with it.

\textit{Step4}. Create a VM cluster and add to the datacenter. Here we create a simple 4 VM serverless cluster with 4 vCPUs and 3 GB of memory each.

\textit{Step5}. Create a new instance of the loadBalancer, which determines the execution flow for incoming function requests.

\textit{Step6}. Create the request workload. We consider an example with user requests arriving at the ServerlessController for a single function deployed in the cluster. First the ServerlessRequest objects are created and submitted to the controller. Once the simulation starts, SimEvents are created for each of them at their particular arrival times when the controller entity starts its functionality. 

\begin{figure}[!ht]
	\centering 
	\includegraphics[width=\linewidth, height=2.8cm]{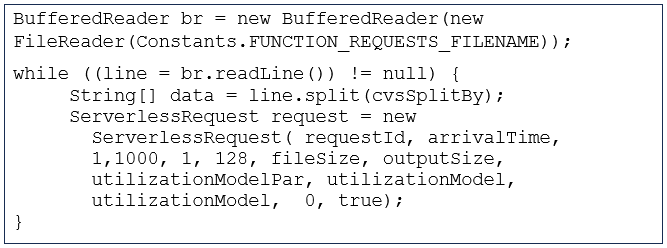}
	\caption{Sample code snippet for creating a function request workload}
	\label{fig:Code1}
\end{figure}

\textit{Step7}. Determine the load balancing policy. For this example we consider a serverless architecture without container concurrency and creating a new container for every new request (this is achieved by enabling 'scale per request' in the configuration file). Thus a container creation request is sent to the ServerlessDataCenter with every new request. 

\begin{figure}[!ht]
	\centering 
	\includegraphics[width=\linewidth, height=2.7cm]{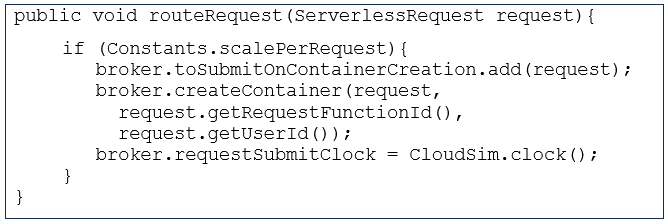}
	\caption{Sample code snippet for routing the requests}
	\label{fig:Code2}
\end{figure}

\textit{Step 8}. Determine the function scheduling policy. We use the default implemented RR policy for this example. Accordingly, every new container created is scheduled on the created VMs in a RR manner. 

\textit{Step 9}. Finally, start the simulation when all simulation parameters are set as needed. Once the execution of all the requests is complete, stop the simulation. If monitoring is enabled, a summary of the VM resource usage details and the request execution details during simulation will be printed on the console.

Note that in the scenario considered here, we did not have to specify a policy for function auto scaling since we followed 'scale per request' where every container accommodates only one request during its life time and the cluster maintains no idling instance pools. Users are encouraged to enable and explore horizontal as well as vertical function scaling by defining their own logic for triggering the scaling functionality and implementing custom policies for both horizontal and vetical scaling.

\section{Performance Evaluation}

In this section we present the simulation of two serverless computing use cases which use the developed components of CloudSimSC. These examples demonstrate the usability of the presence of a generalized simulation environment for serverless computing, in order to evaluate the feasibility of potential resource management techniques prior to deployment in a practical testbed. The first scenario is designed to execute serverless functions following the architecture seen in commercial serverless platforms with a function scheduling algorithm which aims to optimize the usage of cluster VM resources. The second scenario illustrates the execution flow seen in the majority of open-source frameworks, along with the enabled auto-scaling feature.  

\subsection{Case Study 1: Request load balancing and function scheduling}

\begin{figure*}[!tb]
    \centering
    \subfigure[Average Request Response Time (ARRT)]{\includegraphics[width=.4\textwidth, height=4cm]{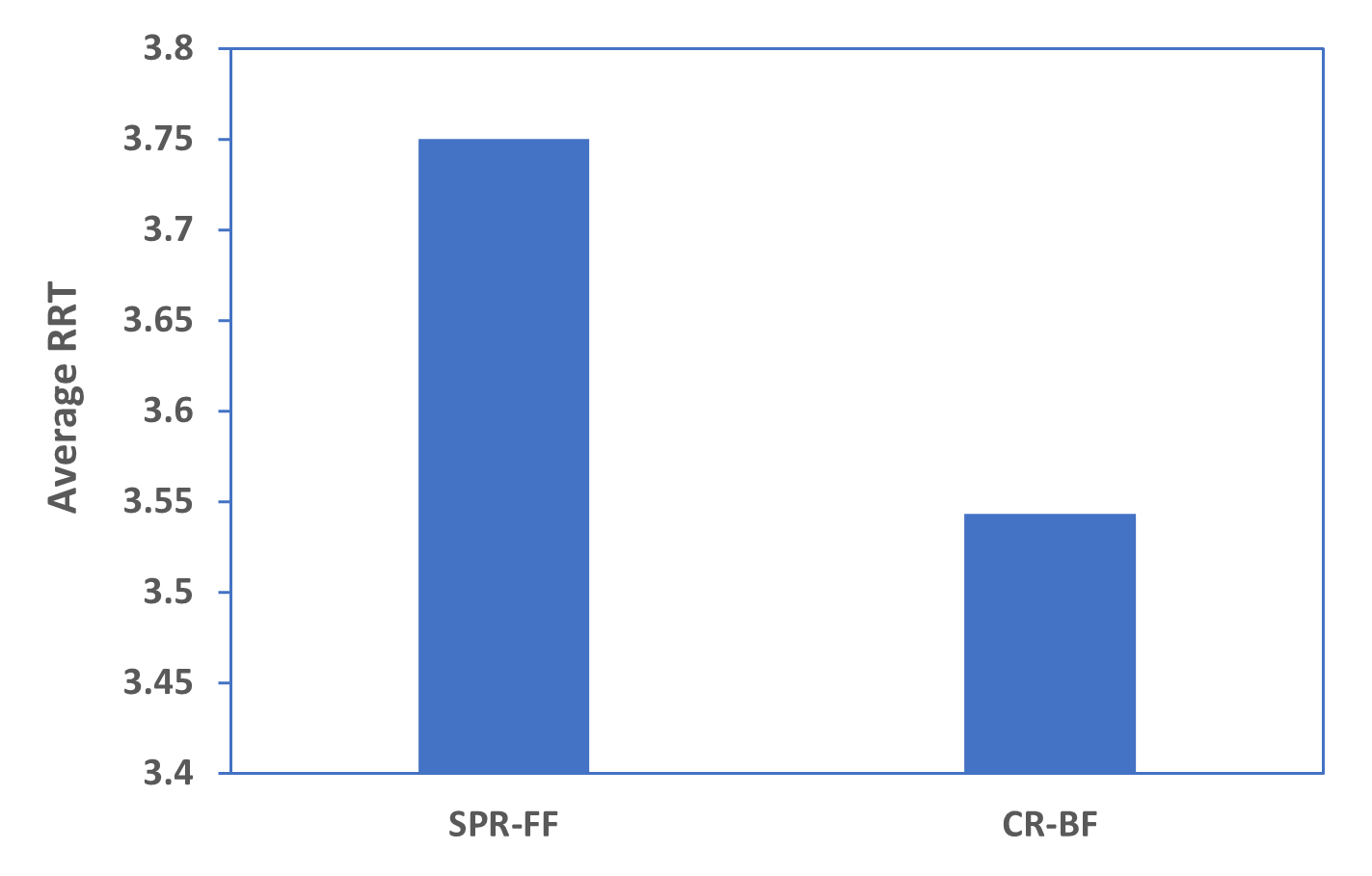}}
    \subfigure[Average VM Utilization]{\includegraphics[width=.4\textwidth, height=4cm]{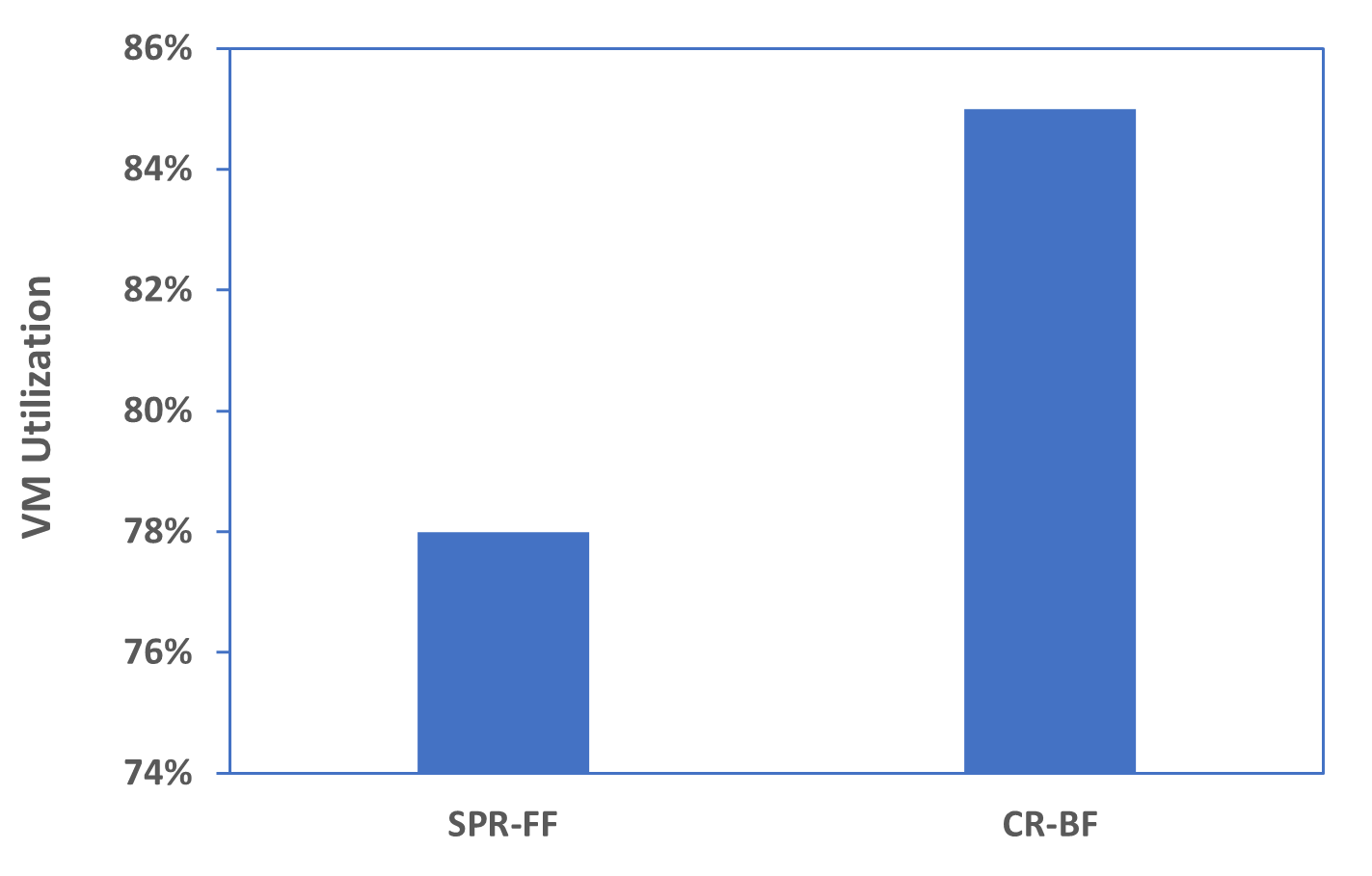}} \\
    \caption{Case Study 1: Comparison of the Average RRT and Average VM Utilization}
    \label{fig:evaluation1}
\end{figure*}

Here we consider a scenario where every function request is accommodated on a separate container, which is able to execute only a single request at a time. For every new request, either a new container (scale per request) is created or an idle warm instance is used. Once a new container is initiated with the function code and dependencies installed, a decision has to be made on selecting a VM to deploy it. This is known as the function scheduling problem which is explored in many research works \cite{kaffes2019centralized}, \cite{suresh2020ensure}, \cite{mampage2021deadline}. The decision on selecting a node for execution affects the efficiency of resource usage, while the reuse of used instances saves up on request scheduling delays. 

\subsubsection{Simulation environment}

We simulate a cluster of 20 homogeneous VMs, each with 4 vCPUs and 3GB of memory and having the clock speed of Intel E5-2666, a configuration seen within AWS Lambda infrastructure \cite{wang2018peeking}. We consider an average startup time for a container to be 500ms. A workload with real-world arrival patterns is created by using trace snippets from Wikipedia \cite{Wikipedi33:online}. These are combined with container size and execution time data from Azure functions data set \cite{shahrad2020serverless}. The created workload consists of traffic arriving for 8 single request applications for a period of one hour, with a peak load of 16x requests per second per application.

\subsubsection{Comparing policies}
As per the generalized architectural features introduced in CloudSimSC, a user could choose to follow the overall request execution flow they wish to simulate. Within the chosen process flow, they could implement and compare the load balancing and instance scheduling policies based on their optimization objectives. Accordingly we conduct experiments using the below policies and compare their results.\\
\textbf{SPR-FF} - A new container is created for every new request. The container is deployed on the first VM which satisfies the resource requirements, out of the set of VMs \\
\noindent
\textbf{CR-BF} - Containers are retained in an idle state after request executions for a set time duration. New requests either choose the first available idle container of that function type or create a new instance and get scheduled on VMs following a bin packing policy of best-fit, where the VMs with higher resource utilization get packed first.

\begin{figure*}[!tb]
    \centering
    \subfigure[Average Request Response Time (ARRT)]{\includegraphics[width=.4\textwidth, height=4cm]{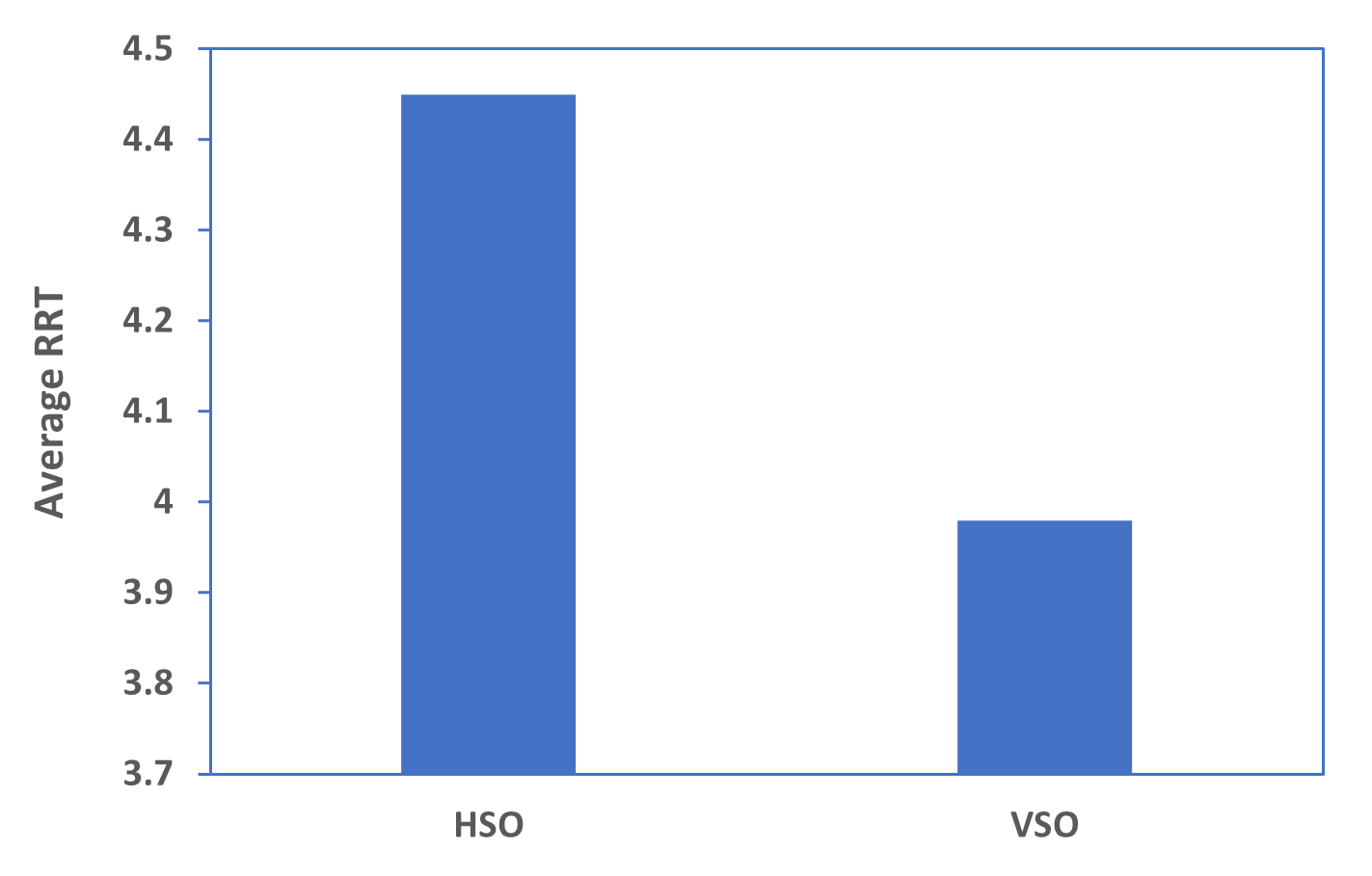}}
    \subfigure[Average VM Utilization]{\includegraphics[width=.4\textwidth, height=4cm]{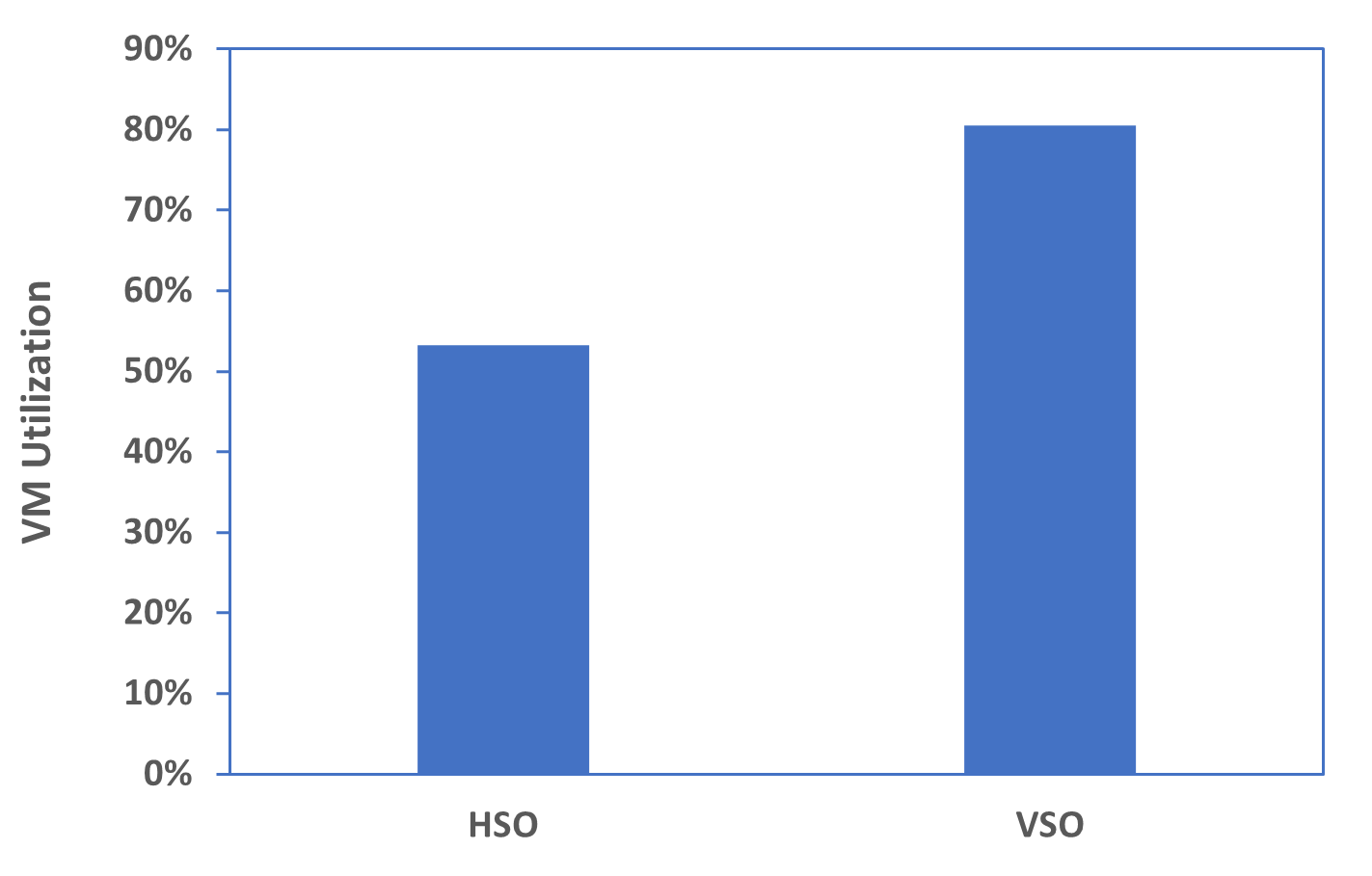}} \\
    \caption{Case Study 2: Comparison of the Average RRT and Average VM Utilization}
    \label{fig:evaluation2}
\end{figure*}

\subsubsection{Results}
The lower average RRT for CR-BF in graph \ref{fig:evaluation1}(a) indicates that the response time latency could be reduced by retaining used containers for future use, which reduces the frequency of cold starts. In contrast, creating a new container for each new request adds a delay factor for each execution. In graph \ref{fig:evaluation1}(b), it is seen that the CR-BF policy increases the resource utilization too. This is partially due to the selection of the VMs with highest utilization for container scheduling, in addition to the retention of used containers. 

\subsection{Case Study 2: Function Auto-scaling}

In the second scenario we explore the serverless architecture followed in the majority of kubernetes based open-source platforms. Here a single function instances accommodates multiple requests at a time based on its resource capacities. Since resources are not scaled per request under this scenario, we implement an auto-scaling policy for expanding resources allocated for a particular application, based on the traffic demand levels. 

\subsubsection{Simulation environment}
Our simulation environment consists of 12 homogenous VMs with the same configuration as in scenario 1. Traces from Azure functions data set are extracted for workload creation for 8 applications receiving requests simultaneously. We maintain a maximum cpu and memory allocated to a function instance at 1 vCPU and 3 GB respectively.

\subsubsection{Comparing policies}
CloudSimSC allows the implementation of both horizontal and vertical scaling policies. Here we implement and compare a few policies, evaluating their effectiveness in terms of function performance and provider resource efficiency. 
.\\
\noindent
\textbf{HSO} - Whenever the average cpu utilization across instances of an application exceeds beyond a set threshold, a set of new replicas are created in order to maintain the utilization at the required level. Down scaling of idling instances is too done following the same logic.\\
\noindent
\textbf{VSO} - Upon reaching step resource utilization thresholds, the cpu and memory capacity of a function instance is incremented or decremented by a pre-defined step value as allowed by the capacity of the VM hosting the instance. \\

\subsubsection{Results}
A lower ARRT is seen in graph \ref{fig:evaluation2}(a) for VSO. This is due to the lack of new resource creation time for vertical scaling, compared to the time spent for new replica creation that is involved with horizontal scaling. Since vertical scaling encourages utilizing the already active VM resources more, it leads to a higher VM utilization level on average, as seen in graph \ref{fig:evaluation2}(b).

\section{Conclusions and Future Work}
Simulation based experiments are instrumental in the investigation of the viability of cloud resource management techniques, specially when dealing with testbed environments which are less scalable and too costly for testing purposes. Serverless computing is a novel computing paradigm which has attracted much interest in the cloud computing community recently. To this end, in this paper, we presented CloudSimSC, a toolkit for modeling a serverless computing environment, developed as an extension to the CloudSim simulator. In contrast to a number of other simulators in the area, CloudSimSC holds the flexibility in following multiple serverless platform architectures as per the user requirement, in addition to the ability to incorporate custom load balancing, scheduling and scaling policies. The case studies discussed in the paper demonstrate the usability of this simulation environment for testing and evaluating various resource management techniques with different optimization objectives. 

As part of future work, we plan to extend CloudSimSC to support multi-function applications with complex Directed Acyclic Graph (DAG) based workflow structures. 
\\ \\
\noindent
\textbf{Software availability}: The source code of the CloudSimSC toolkit is accessible from:
https://github.com/Cloudslab/CloudSimSC

\bibliographystyle{./bibliography/IEEEtran}
\bibliography{./bibliography/Serverless.bib}

\end{document}